# Radiosensitization beyond DNA damage: Monte Carlo simulations of realistic nanomaterial biodistributions

**Christian Velten[1,2], Brett Bell[2], and Wolfgang A. Tomé[1,2]**
[1] Montefiore Medical Center, Bronx, NY, USA
[2] Albert Einstein College of Medicine, Bronx, NY, USA

E-mail: cvelten@montefiore.org

**Abstract.** We investigated cellular distribution of a tumor-specific gadolinium chelate in 4T1 and U87 cancer cells with the goal to generate more realistic geometries for Monte Carlo simulations of radiation interaction with nanoparticles in cells. Cells were exposed to the agent in-vitro for 30 minutes to 72 hours before being fixed and imaged using transmission electron microscopy. Initially, electron-dense areas consistent with gadolinium were observable throughout the cytoplasm. At six hours those areas were restricted to endosomes and at 24 hours or longer electron dense areas were only found in lysosomes. Lysosomes were on average larger in 4T1 cells, which were exposed to 1 mM concentration compared to U87 cells, exposed to 1 µM. Based on this information we built an extended cell model for Monte Carlo simulations that includes lysosomes with discrete nanoparticle enclaves in addition to mitochondria and the nucleus.

## 1. Introduction

Radiation therapy is an important tool in the fight against cancer and has undergone far-reaching technological improvements. Its utility, however, is usually limited by the rate and severity of normal tissue toxicities. Further widening the therapeutic window beyond what is currently possible with state of the art technology, the difference between tumor control and normal tissue complication probabilities, could further improve local control while reducing normal tissue toxicities. Nanomaterials, especially those with high atomic number, have long been studied as potential radiosensitizers. Several in-vitro and in-vivo studies observed decreased cell survival with nanomaterials but did not find significant increases in DNA damage [1–3], while others saw an increase in DNA damage at doses well in excess of 50 Gy [4]. This suggests that non-DNA mediated cell death pathways involving the mitochondria and lysosomes [5–8] may become important in the presence of metallic nanomaterials, especially at clinically relevant doses.

Here, we investigated the biodistribution of a tumor-specific alkylphosphocholine gadolinium chelate (Gd-NM600) in cancer cells [9–13], with the aim to generate more realistic geometries for Monte Carlo particle transport simulations. These geometries will include both sub-cellular structures and discrete nanoparticles to properly handle the stochastic nature of nanoparticle interactions [14,15].

## 2. Materials and Methods

### 2.1. Cell Cultures and Transmission Electron Microscopy

4T1 breast cancer and U87 glioma cell lines were cultured in DMEM containing 10% fetal bovine serum, 100 U/mL penicillin, and 100 µg/mL streptomycin. Cells were concurrently plated in 60 mm polystyrene dishes and allowed to adhere overnight. 4T1 cells were exposed to 1 mM Gd-NM600 for 24 h and 72 h with controls at both times. U87 cells were exposed to 1 µM Gd-NM600 for 0.5 h, 6 h, and 24 h with one control at 24 h. Cells were rinsed with serum-free media, fixed in 2% paraformaldehyde and 2.5% glutaraldehyde for two hours, before being transferred to 0.1 M sodium phosphate buffer, and stored at 4°C until preparation for transmission electron microscopy (TEM). Samples were then postfixed with 1% osmium tetroxide followed by 2% uranyl acetate, dehydrated through a graded series of ethanol and embedded in LX112 resin (LADD Research Industries, Burlington VT). Ultrathin (80 nm) sections were cut on a Leica EM Ultracut UC7, optionally stained with uranyl acetate followed by lead citrate and viewed on a JEOL 1400 Plus transmission electron microscope at 120 kV. An extended cell model was created as a geometry component extension for TOPAS [16,17] based on the existing TOPAS-nBio spherical cell models [18].

## 3. Results

Mitochondria and lysosomes were counted from ten different cells in each cell line's control group, while their dimensions were measured within a single cell each. Average dimensions of mitochondria and lysosomes from 4T1 and U87 cells are tabulated in Table 1.

Table 1. Average dimensions of mitochondria and lysosomes obtained from several 4T1 and U87 cells, where $\langle a \rangle$ and $\langle b \rangle$ are the long and short axes of a measured organelle.

|  |  | n | $\langle a \rangle$ / μm | $\sigma$ / μm | $\langle b \rangle$ / μm | $\sigma$ / μm |
|---|---|---|---|---|---|---|
| **4T1** | Mitochondria | 17 | 0.50 | 0.18 | 0.36 | 0.11 |
|  | Lysosomes | 27 | 0.87 | 0.37 | 0.78 | 0.37 |
| **U87** | Mitochondria | 24 | 0.41 | 0.13 | 0.28 | 0.06 |
|  | Lysosomes | 23 | 0.57 | 0.19 | 0.45 | 0.19 |

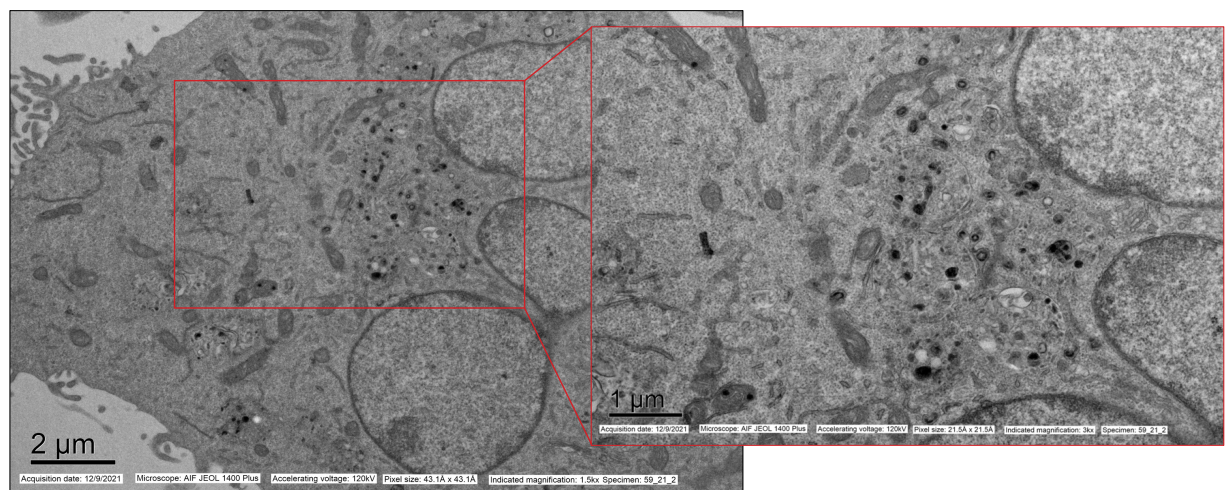

**Figure 1.** Transmission electron micrograph of 4T1 cancer cells exposed to 1 mM Gd-NM600 for 72 h. Sections were unstained, i.e. no use of lead citrate or uranyl acetate. Electron dense regions consistent with presence of gadolinium can be identified within lysosomes, highlighted in the insert.

Electron dense areas consistent with the presence of gadolinium were observed in all exposed samples. At less than 24 h electron dense areas were present in the cytoplasm and in proximity to endosomes. After 24 h or more of exposure, electron dense areas were primarily found within the enzyme complexes of lysosomes for both cell lines (Figures 1 & 2). Electron dense regions in 4T1 cells were not considerably more dense or prevalent than those in U87 cells, despite exposure to 1000x higher Gd-NM600 concentration. However, 4T1 cells exposed to 1 mM for 72 h were less dense than their 72 h control group, without observable remnants of cell death.

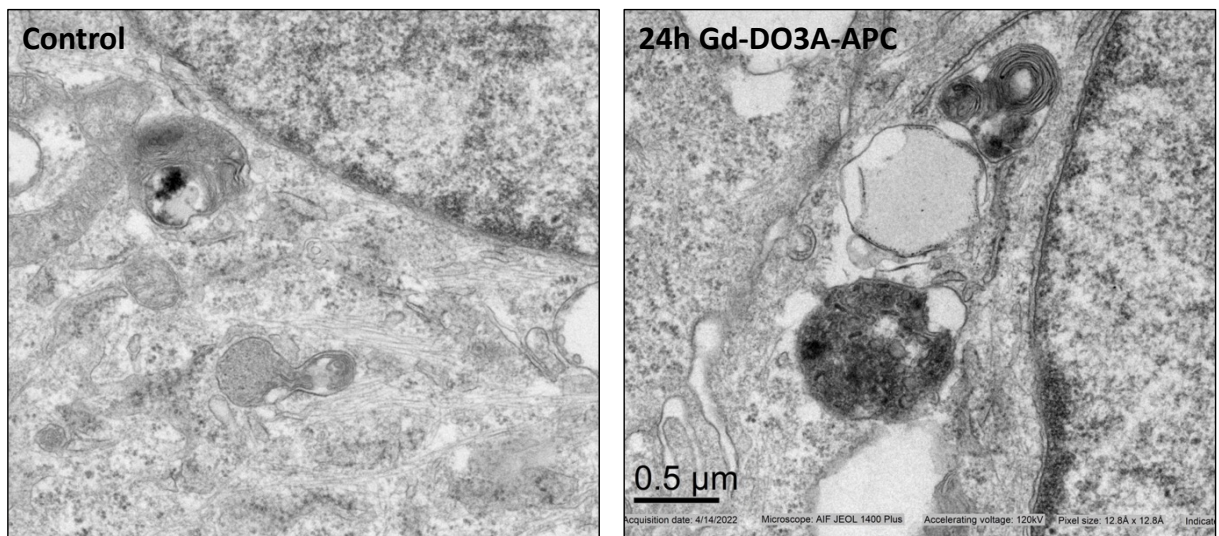

**Figure 2.** U87 glioma cells without (left panel) and with 24 h exposure to Gd-DO3A-APC (right panel), scale applies to both panels. Substantial accumulation of electron dense material consistent with presence of gadolinium can be seen in the lysosomes, with negligible electron density in the control. Micrographs were stained using lead citrate and uranyl acetate.

*3.1. Monte Carlo Simulation Setup*

The spherical cell model included with TOPAS-nBio was extended to include lysosomes in addition to mitochondria and a spherical nucleus. Discrete nanoparticles spheres were added to the lysosomes (Figure 3). While the NP were found to accumulate within the enzyme complexes, these were not independently modeled due to the wide ranges in size and dimension and are only shown as an example in Figure 3. The number of NP per lysosome was calculated based on the total cellular NP mass and the (average) number of lysosomes. The average dimension of mitochondria and lysosomes was obtained from a sample of cells imaged using TEM (Table 1), while the average number of approximately 250-500 was estimated from literature values [18] due to their large variability within even one cell line due to differences in cell cycle and cellular environment.

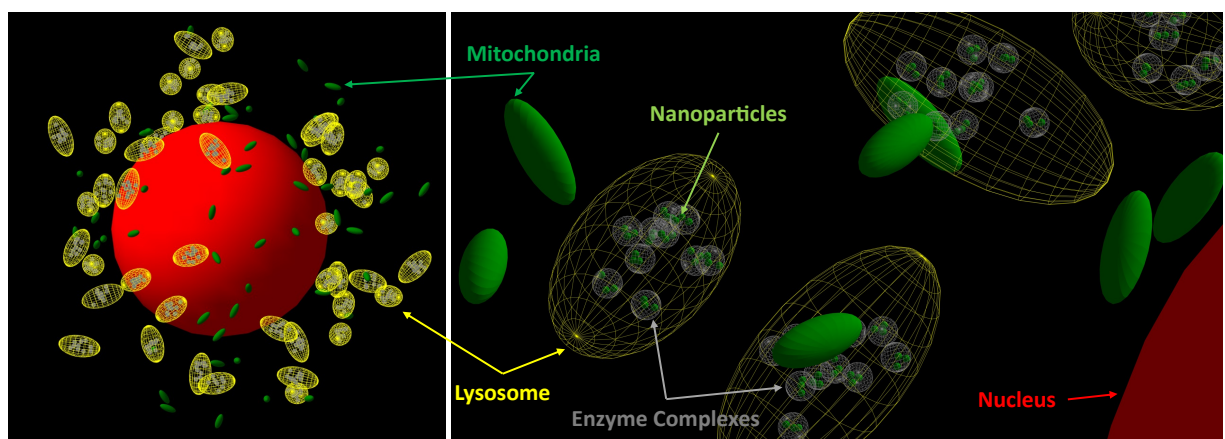

**Figure 3.** TOPAS/Geant4 cell model including a spherical nucleus, ellipsoidal mitochondria and lysosomes with nanoparticle within enzyme complexes.

4. **Discussion and Conclusion**

The development of electron dense areas with time indicates that Gd-NM600 is initially distributed throughout the cytoplasm before aggregating in endosomes, which then fuse with lysosomes. With no clear difference in electron dense areas between 4T1 cells exposed to 1 mM and U87 cells exposed to 1 μM, this may indicate saturable uptake of Gd-NM600, while differences in uptake and retention rates between different cell lines may be equally possible. High concentrations of 1 mM are likely not achievable in the clinical setting and the observed proliferation inhibition is thus negligible. While mitochondria were similar in size between the two cell lines, lysosomes were larger in 4T1 cells exposed to higher concentrations, though only within one standard deviation.

Accumulation and retention of this agent in lysosomes could lead to radiosensitization through the local increase of reactive oxygen species that can trigger lysosome and mitochondria dependent cell death. Further studies will aim to quantify the accumulation and radiation effects.

Finally, while this study used a gadolinium chelate of NM600, combinations with other metals including radioactive isotopes are equally feasible and expected to have comparable uptake and distribution behavior.

## 5. Acknowledgements

Jamey Weichert at the University of Wisconsin for providing the Gd-NM600. Chandan Guha lab at Albert Einstein College of Medicine for providing the cell lines. Analytical Imaging Facility at Albert Einstein College of Medicine for TEM, supported by NIH P30CA013330 and NIH SIG 1S10OD016214-01A1.

## References

[1] Jain S et al. Cell-Specific Radiosensitization by Gold Nanoparticles at Megavoltage Radiation Energies. Int J Radiat Oncol Biol Phys 2011; 79: 531–9. doi:10.1016/j.ijrobp.2010.08.044.
[2] Kim J-K et al. Enhanced proton treatment in mouse tumors through proton irradiated nanoradiator effects on metallic nanoparticles. Phys Med Biol 2012; 57: 8309–23. doi:10.1088/0031-9155/57/24/8309.
[3] Štefančíková L et al. Effect of gadolinium-based nanoparticles on nuclear DNA damage and repair in glioblastoma tumor cells. J Nanobiotechnology 2016; 14: 63. doi:10.1186/s12951-016-0215-8.
[4] Schlatholter et al. Improving proton therapy by metal-containing nanoparticles: nanoscale insights. Int J Nanomedicine 2016; 11: 1549. doi:10.2147/IJN.S99410.
[5] Boya P, Kroemer G. Lysosomal membrane permeabilization in cell death. Oncogene 2008; 27: 6434–51. doi:10.1038/onc.2008.310.
[6] Galluzzi L et al. Molecular mechanisms of cell death: recommendations of the Nomenclature Committee on Cell Death 2018. Cell Death Differ 2018; 25: 486–541. doi:10.1038/s41418-017-0012-4.
[7] Wang F et al. Lysosomal membrane permeabilization and cell death. Traffic 2018; 19: 918–31. doi:10.1111/tra.12613.
[8] Luke CJ et al. Lysoptosis is an evolutionarily conserved cell death pathway moderated by intracellular serpins. Commun Biol 2022; 5: 47. doi:10.1038/s42003-021-02953-x.
[9] Weichert JP et al. Alkylphosphocholine Analogs for Broad-Spectrum Cancer Imaging and Therapy. Sci Transl Med 2014; 6: 240ra75-240ra75. doi:10.1126/scitranslmed.3007646.
[10] Hernandez R et al. 90Y-NM600 targeted radionuclide therapy induces immunologic memory in syngeneic models of T-cell Non-Hodgkin's Lymphoma. Commun Biol 2019; 2: 79. doi:10.1038/s42003-019-0327-4.
[11] Grudzinski JJ et al. Preclinical Characterization of 86/90 Y-NM600 in a Variety of Murine and Human Cancer Tumor Models. Journal of Nuclear Medicine 2019; 60: 1622–8. doi:10.2967/jnumed.118.224808.
[12] Hernandez R et al. 177 Lu-NM600 Targeted Radionuclide Therapy Extends Survival in Syngeneic Murine Models of Triple-Negative Breast Cancer. Journal of Nuclear Medicine 2020; 61: 1187–94. doi:10.2967/jnumed.119.236265.
[13] Zhang RR et al. Next-Generation Cancer Magnetic Resonance Imaging with Tumor-Targeted Alkylphosphocholine Metal Analogs. Invest Radiol 2022; 57: 655–63. doi:10.1097/RLI.0000000000000893.
[14] Hahn MB, Zutta Villate JM. Combined cell and nanoparticle models for TOPAS to study radiation dose enhancement in cell organelles. Sci Rep 2021; 11: 1–10. doi:10.1038/s41598-021-85964-2.
[15] Rabus H et al. Intercomparison of Monte Carlo calculated dose enhancement ratios for gold nanoparticles irradiated by X-rays: Assessing the uncertainty and correct methodology for extended beams. Physica Medica 2021; 84: 241–53. doi:10.1016/j.ejmp.2021.03.005.
[16] Perl J et al. TOPAS: An innovative proton Monte Carlo platform for research and clinical applications. Med Phys 2012; 39: 6818–37. doi:10.1118/1.4758060.
[17] Faddegon B et al. The TOPAS tool for particle simulation, a Monte Carlo simulation tool for physics, biology and clinical research. Physica Medica 2020; 72: 114–21. doi:10.1016/j.ejmp.2020.03.019.
[18] Schuemann J et al. TOPAS-nBio: An Extension to the TOPAS Simulation Toolkit for Cellular and Sub-cellular Radiobiology. Radiat Res 2018; 191: 125. doi:10.1667/RR15226.1.